\documentclass[aps,prl,twocolumn,showpacs,groupedaddress,bibnotes]{revtex4}

\usepackage{graphicx}
\usepackage{bm}

\graphicspath{{pict/}{}}
\newcommand\pictc[5]{\begin{figure}
            \centerline{\vspace{0mm}
\includegraphics[width=#1\columnwidth,height=0.7\textheight,keepaspectratio]{#3}}
            \protect\caption{\protect\label{fig:#4} #5}\vspace{-0mm}
                    \end{figure}            }
                      
\newcommand\pict[4][1.0]{\pictc{#1}{!tb}{#2}{#3}{#4}}

\newcommand\pictcWide[5]{\begin{figure*}
            \centerline{\vspace{0mm}
\includegraphics[width=#1\textwidth,height=0.7\textheight,keepaspectratio]{#3}}
            \protect\caption{\protect\label{fig:#4} #5}\vspace{-0mm}
                    \end{figure*}            }
                      
\newcommand\pictWide[4][1.0]{\pictcWide{#1}{!tb}{#2}{#3}{#4}}

\newcommand\rpict[1]{\ref{fig:#1}}

\newcounter{Fig}

\begin{document}
\begin{sloppy}

\title{Diffraction-managed solitons and nonlinear beam diffusion\\
in modulated arrays of optical waveguides}


\author{Alexander Szameit$^1$, Ivan L. Garanovich$^2$, Matthias Heinrich$^1$, Alexander Minovich$^2$, Felix Dreisow$^1$, Andrey A. Sukhorukov$^2$, Thomas~Pertsch$^1$, Dragomir N. Neshev$^2$, Stefan Nolte$^1$, Wieslaw Krolikowski$^2$, Andreas T\"{u}nnermann$^1$, Arnan~Mitchell$^3$, and Yuri S. Kivshar$^2$}

\affiliation{
$^1$Institute of Applied Physics, Friedrich-Schiller-University Jena,
\mbox{Max-Wien-Platz 1, 07743 Jena, Germany}\\
$^2$Centre for Ultra-high bandwidth Devices for Optical Systems (CUDOS),\\
Nonlinear Physics Centre and Laser Physics Centre, RSPhysSE, Australian National University, Canberra, Australia\\
$^3$School of Electrical and Computer Engineering, RMIT University, Melbourne, Australia
}

\begin{abstract}
We study propagation of light in nonlinear diffraction-managed photonic lattices created with arrays of periodically-curved coupled optical waveguides which were fabricated using femtosecond laser writing in silica glass, and titanium indiffusion in LiNbO$_3$ crystals. We identify different regimes of the nonlinear propagation of light beams depending on the input power, and present {\em the first experimental observation of diffraction-managed solitons}, which are formed as a result of the interplay between the engineered beam diffraction and nonlinear self-focusing or defocusing. We observe that in self-collimating structures where linear diffraction is suppressed, a novel regime of {\em nonlinear beam diffusion} takes place at the intermediate powers before the lattice soliton is formed at higher powers.
\end{abstract}

\pacs{42.65.Jx, 42.82.Et, 42.65.Tg}


\maketitle

Propagation of light in dielectric media with a periodically-varying refractive index is known to demonstrate many novel features in both linear and nonlinear regimes~\cite{Christodoulides:2003-817:NAT}. Over the recent years, different types of periodic photonic structures, including arrays of evanescently coupled optical waveguides, optically-induced lattices in photorefractive materials, and photonic crystals, have been employed to engineer and control the fundamental properties of light propagation. In particular, the idea to control the light spreading through diffraction management~\cite{Eisenberg:2000-1863:PRL} has attracted a special attention. It was shown that both magnitude and sign of the beam diffraction can be controlled in periodic photonic structures. For example, diffraction can be made {\em negative} allowing for focusing of diverging beams~\cite{Pertsch:2002-93901:PRL}, or it can even be suppressed leading to the self-collimation effect when the beam width does not change over hundreds of the free-space diffraction lengths~\cite{Rakich:2006-93:NAMT}. Self-collimation of light beams was also demonstrated in arrays of periodically curved optical 
waveguides~\cite{Longhi:2006-243901:PRL, Longhi:2005-2137:OL, Iyer:2007-3212:OE}. Recently, the concept of the broadband diffraction management of supercontinuum light has been introduced for the modulated waveguide arrays with special axes bending profiles~\cite{Garanovich:2006-066609:PRE},
suggesting novel opportunities for the manipulation of polychromatic light beams and patterns.

The combination of tailored diffraction characteristics and light self-action opens new possibilities for the power-controlled beam shaping and switching in nonlinear photonic structures, enabling dynamical tunability of photonic devices. Various schemes for active beam control based on the special properties of narrow self-localized beams in straight waveguide arrays, called discrete spatial solitons, have been suggested and demonstrated~\cite{Christodoulides:1988-794:OL, Eisenberg:1998-3383:PRL, Kivshar:2003:OpticalSolitons}. It was suggested theoretically, that solitons can also exist in diffraction-managed lattices~\cite{Ablowitz:2001-254102:PRL, Staliunas:2007-11604:PRA} in the regime when beam exhibits effectively averaged diffraction. 

Recently, self-action of narrow light beams in diffraction-managed periodically curved nonlinear waveguide arrays has been analyzed beyond the applicability of averaging procedures, and novel dynamical regimes have been identified depending on the input power~\cite{Garanovich:2007-9547:OE}. 
In particular, it was predicted that in periodically curved waveguide arrays [such as sketched in Fig.~\rpict{fig1_Theory_34um}(f)] with canceled effective diffraction transition from the regime of self-collimation, at low powers, to that of the discrete self-trapping and formation of the diffraction-managed lattice solitons, at high powers, should occur through the intermediate regime of the nonlinear beam diffusion, as shown in Fig.~\rpict{fig1_Theory_34um}(f)-(k). This is in a sharp contrast to the monotonous beam self-focusing which takes place in straight waveguide arrays as the input power increases, see Fig.~\rpict{fig1_Theory_34um}(a)-(e). However, nonlinear beam self-action in diffraction-managed lattices have not yet been studied experimentally, and no type of diffraction-managed solitons have been observed in experiment in photonic lattices so far.

\pictWide[1.0]{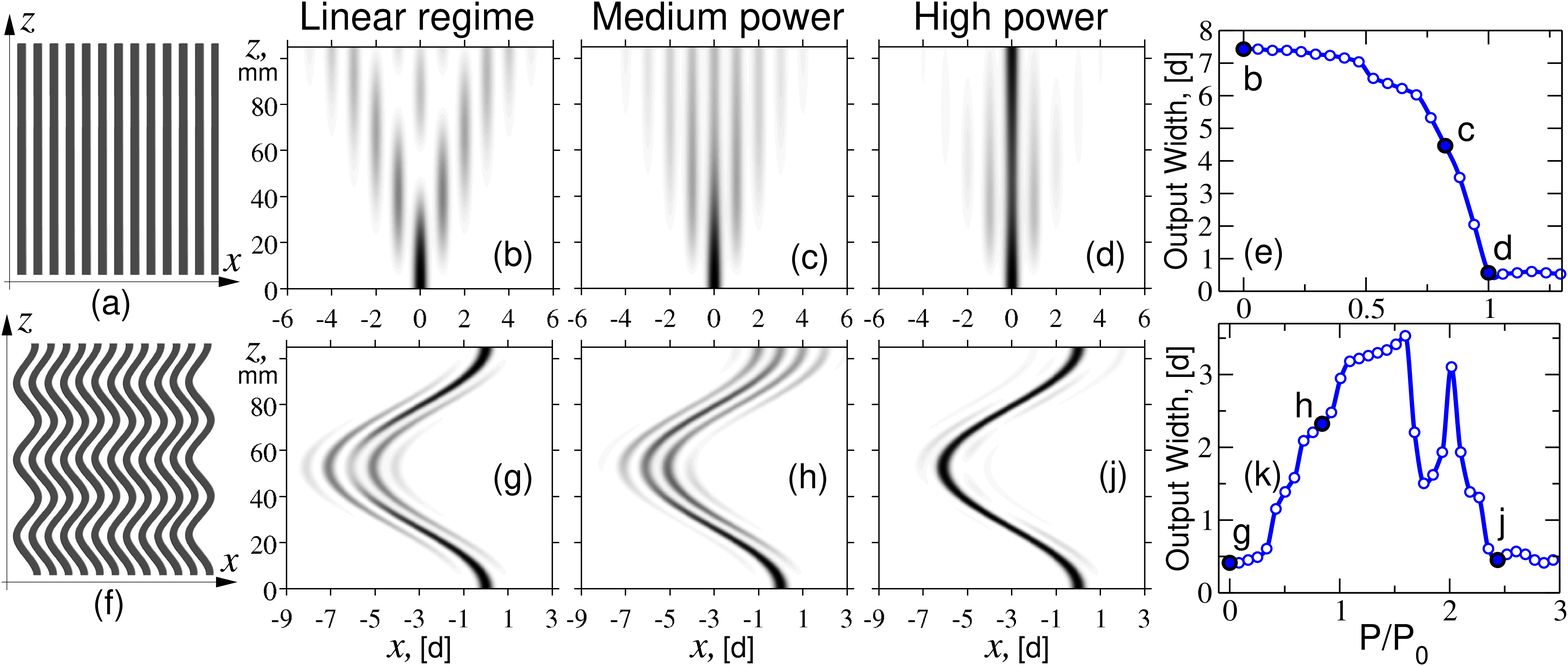}{fig1_Theory_34um}{
Numerical simulations of beam propagation in straight [top] and periodically-curved [bottom] nonlinear waveguide arrays.
(b,c,d)~Discrete diffraction, beam self-focusing, and lattice soliton in (a) straight array.
(g,h,j)~Self-collimation, nonlinear beam diffusion, and diffraction-managed soliton in (f) curved array.
(e,k)~Output beam width vs. the input power for straight and curved arrays.
Points 'b', 'c', 'd' and 'g', 'h', 'j' correspond to the input powers in (b), (c), (d) and (g), (h), (j), respectively.
Input power is normalized to the power of the lattice soliton in the straight array.
Coupling strength and propagation distance are the same as 
in Figs.~\rpict{fig2_Experiment_Glass_straight}(a)~and~\rpict{fig3_Experiment_Glass_curved}(a).
}
 
In this work we present, to the best of our knowledge, the {\em first experimental observation} of {\em nonlinear beam diffusion} and formation of {\em diffraction-managed solitons} in diffraction-managed photonic lattices created with arrays of periodically-curved evanescently coupled nonlinear optical waveguides. We have fabricated curved waveguide arrays in fused silica using the femtosecond laser writing technique. Details of the writing procedure, which was performed using a Mira/RegA (Coherent) system,
can be found elsewhere~\cite{Szameit:2007-1579:OE}.  We created waveguide arrays with a sinusoidal axis bending profile of the form
${\rm x_0(}z{\rm)} =  {\rm A} \lbrace \cos\left[2{\rm \pi} z / {\rm L}\right] - 1 \rbrace$, 
where ${\rm x_0(}z{\rm)}$ is the transverse lattice shift as a function of the propagation distance $z$ which determines the lattice modulation, 
A and L are the waveguide axes bending amplitude and period, respectively.
When the bending amplitude A is such that $\rm 2 \pi \omega A / L = \xi$, where $\omega$  is the normalized frequency
and ${\rm \xi} \simeq 2.40$ is the first root of the 
Bessel function ${\rm J_0}$~\cite{Garanovich:2007-9547:OE, Longhi:2006-243901:PRL}, the effective beam diffraction
is canceled after the propagation over each bending period, and the beam experiences periodic self-collimation with the period L,
in close analogy to the dynamical localization of charged particles in ac electric fileds~\cite{Dunlap:1986-3625:PRB}.
Normalized frequency is $\rm \omega = 2 \pi n_0 d / \lambda$, where ${\rm \lambda}$ is the vacuum wavelength,
${\rm n_0}$ is the average refractive index of the medium, and d is the spacing between the centers of the adjacent waveguides.
In our experiments, we used samples with four different waveguide spacings, d
$=34$~$\rm \mu$m, $36$~$\rm \mu$m, $38$~$\rm \mu$m and $40$~$\rm \mu$m. The waveguide bending amplitudes where chosen to satisfy the self-collimation condition, A$=104$~$\rm \mu$m, $98$~$\rm \mu$m, $93$~$\rm \mu$m, and $88$~$\rm \mu$m, respectively. Each sample consisted of 13 waveguides with elliptical
transverse cross-section with the diameters of approximately 4x13~$\rm \mu$m~\cite{Szameit:2007-1579:OE}, and was L$=105$~mm long.

For each of the curved waveguide arrays we have also fabricated a straight counterpart of the same length, and first characterized linear diffraction and nonlinear focusing in the straight samples. As we increased the input power, we observed monotonous beam self-focusing and formation eventually of a single-site lattice soliton at some threshold power~\cite{Scott:1983-87:PL} [see Fig.~\rpict{fig2_Experiment_Glass_straight}(a)-(c)]. The soliton power was lower for the samples with higher waveguide spacing,
for which the coupling between the waveguides is weaker [see Fig.~\rpict{fig2_Experiment_Glass_straight}(c)
and Fig.~\rpict{fig4_dependencesGlass}(a)].

\pict[1.0]{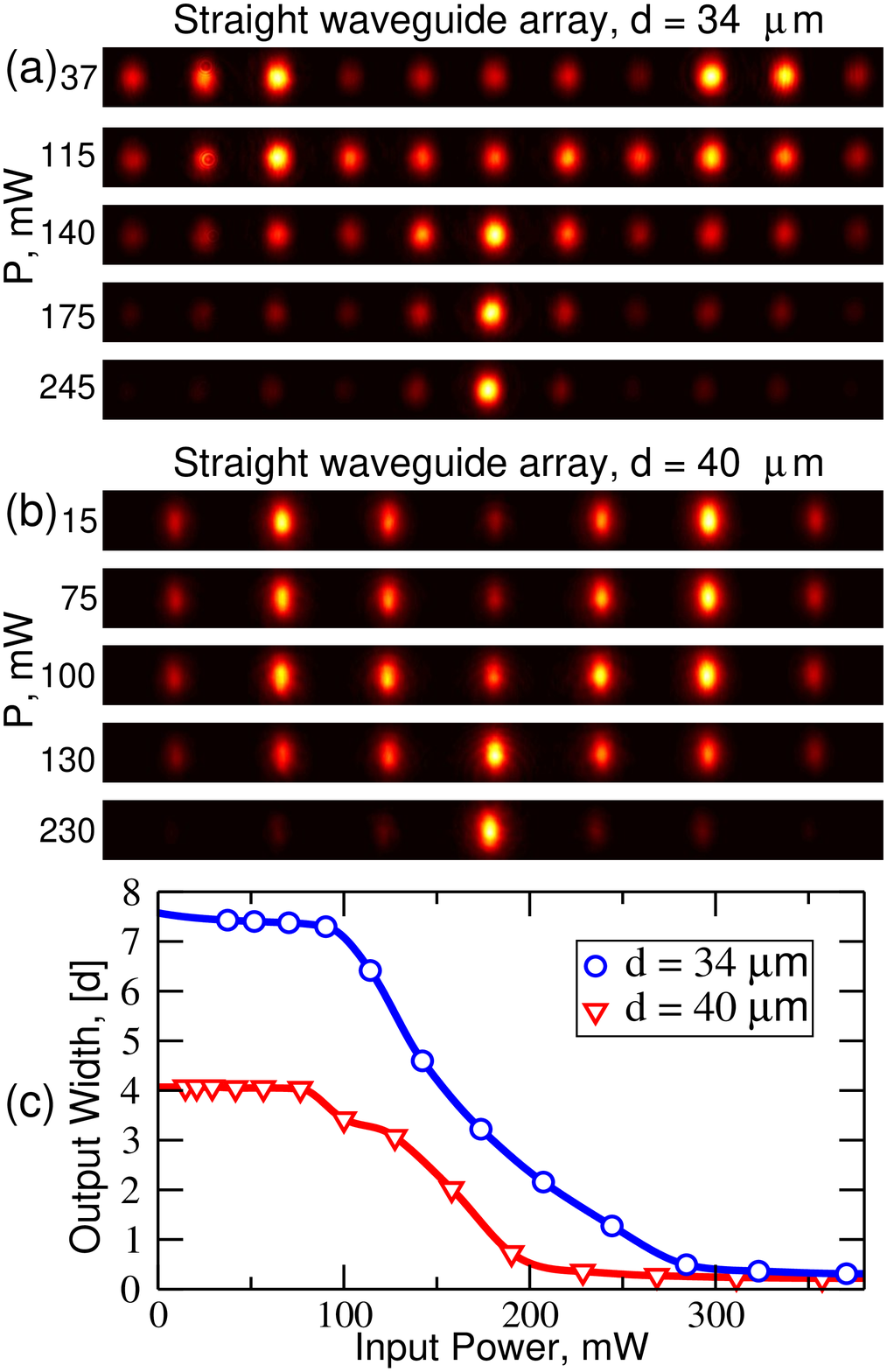}{fig2_Experiment_Glass_straight}{
(a,b)~Output beam profiles as a function of the input power measured in straight fs laser written waveguide arrays 
with waveguide spacing 34~$\rm \mu$m and 40~$\rm \mu$m. (c)~Output beam width vs. the input power. 
Circles correspond to (a), triangles correspond to (b). Wavelength is $\rm \lambda = 800$~nm.
}

Next, we studied propagation of light beams in the curved waveguide arrays for different input laser powers.
At low powers, in the essentially linear propagation regime, we observed dynamical {\em self-collimation} of light in the curved waveguides,
similar to the previous experiments~\cite{Longhi:2006-243901:PRL, Iyer:2007-3212:OE}.
We observed that at the output facet of the arrays all the light was collected back into the same single waveguide in which it was injected initially at the input [see Figs.~\rpict{fig3_Experiment_Glass_curved}(a)~and~(b), top].

\pict[1.0]{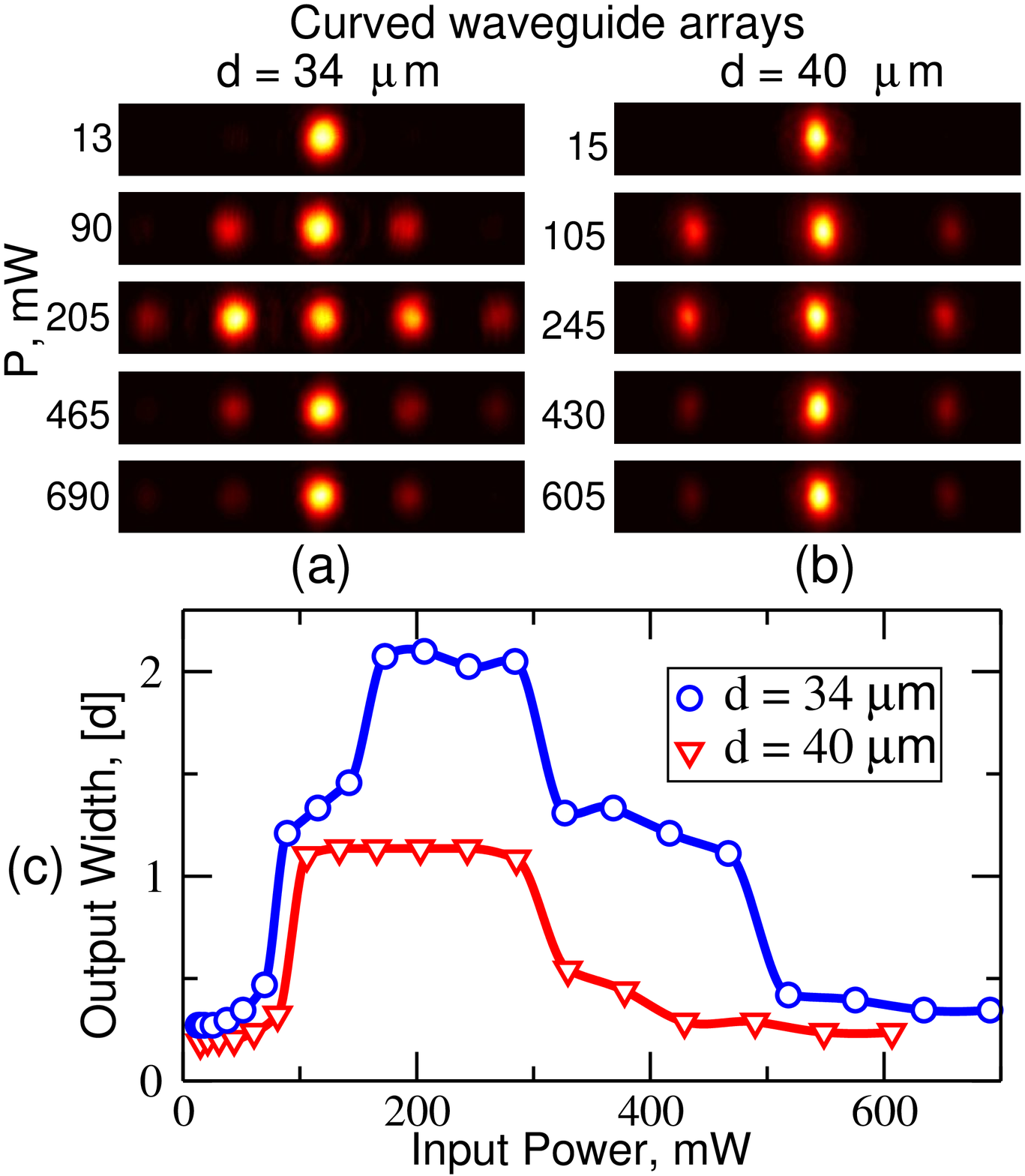}{fig3_Experiment_Glass_curved}{
(a,b)~Output beam profiles as a function of the input power measured in curved fs laser written waveguide arrays 
with waveguide spacing 34~$\rm \mu$m and 40~$\rm \mu$m. (c)~Output beam width vs. the input power. 
Circles correspond to (a), triangles correspond to (b). Wavelength is $\rm \lambda = 800$~nm.
}

When the power of the input beam was increased, we observed {\em nonlinear beam diffusion} in the excellent agreement with
the theoretical predictions~\cite{Garanovich:2007-9547:OE}. The beam experienced self-defocusing and significant broadening
[see Figs.~\rpict{fig3_Experiment_Glass_curved}(a)~and~(b), central images] as nonlinearity destroyed the self-collimation 
condition by changing the refractive index of the waveguide material.
This beam self-defocusing is, however, intrinsically limited because the linear discrete diffraction is fully suppressed in 
the curved waveguide arrays with the bending amplitude which satisfies the self-collimation condition. After propagation over 
some distance the beam broadens and its intensity is reduced accordingly. Therefore, the further beam spreading stops when 
the average beam width achieves a certain value. Such a peculiar nonlinear beam dynamics, which we observed here
for the first time, has no analogies in bulk media~\cite{Kivshar:2003:OpticalSolitons} or in discrete systems~\cite{Christodoulides:2003-817:NAT} analyzed before, where the beam experiences either {\em monotonous} self-focusing or defocusing depending on the type of the nonlinearity 
[compare Fig.~\rpict{fig3_Experiment_Glass_curved} with Fig.~\rpict{fig2_Experiment_Glass_straight}].

\pict[1.0]{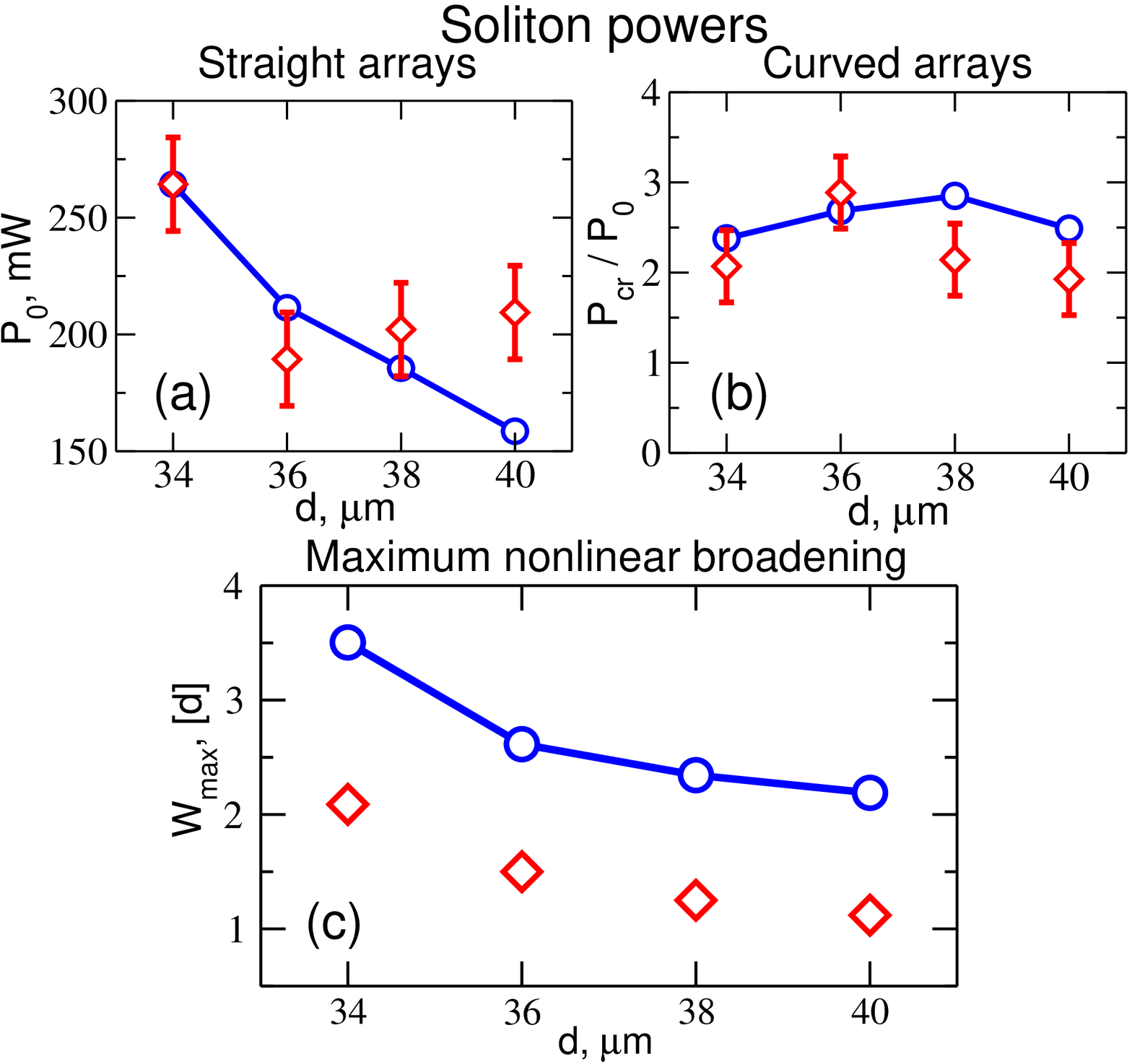}{fig4_dependencesGlass}{
(a,b)~Power P$_0$ of the lattice soliton in straight waveguide arrays, and critical power P$_{cr}$ required for the formation of
diffraction-managed solitons in curved waveguide arrays, as a function of the waveguide spacing d. Diamonds with error bars
and circles connected with solid lines represent
data measured experimentally and calculated theoretically, respectively. In~(b) powers of the diffraction-managed solitons 
in curved arrays are normalized to the powers of the lattice solitons in the straight arrays with the same waveguide spacing.
(c) Diamonds and circles show maximum nonlinear beam broadening measured experimentally and calculated theoretically, respectively,
for the waveguide arrays with different waveguide spacing d. 
}

Then, at higher input powers, nonlinear self-trapping of the beam  to a single lattice site eventually occurred,
what we believe to be the {\em first ever experimental observation} of {\em diffraction-managed solitons}
[see Figs.~\rpict{fig3_Experiment_Glass_curved}(a)~and~(b), bottom]. We observed, in the well agreement with
the theoretical predictions~\cite{Garanovich:2007-9547:OE}, that the power required for the formation of
the diffraction-managed solitons in curved waveguide arrays was more than two times higher than the power of lattice
solitons in exactly the same but straight waveguide arrays 
[compare Fig.~\rpict{fig3_Experiment_Glass_curved}(c) with Fig.~\rpict{fig2_Experiment_Glass_straight}(c),
and also see Figs.~\rpict{fig4_dependencesGlass}(a)~and~(b)]. Error bars
in Figs.~\rpict{fig4_dependencesGlass}(a)~and~(b) represent uncertainty caused by the laser power step which we used
in our measurements. However, there is also an additional uncertainty 
in the determination of the exact power at which the soliton is formed caused
by the slowly dissipating radiation in our relatively short samples, 
which is very difficult to account for quantitatively. In Fig.~\rpict{fig4_dependencesGlass}(c) one can see that 
the beam broadening in the regime of the nonlinear diffusion is less for the waveguide arrays with higher waveguide spacing,
for which the coupling strength between the adjacent waveguides is lower. 
Also, the maximum observed nonlinear beam broadening is less than the theoretical
predictions of the discrete model. This is because???

\pict[1.0]{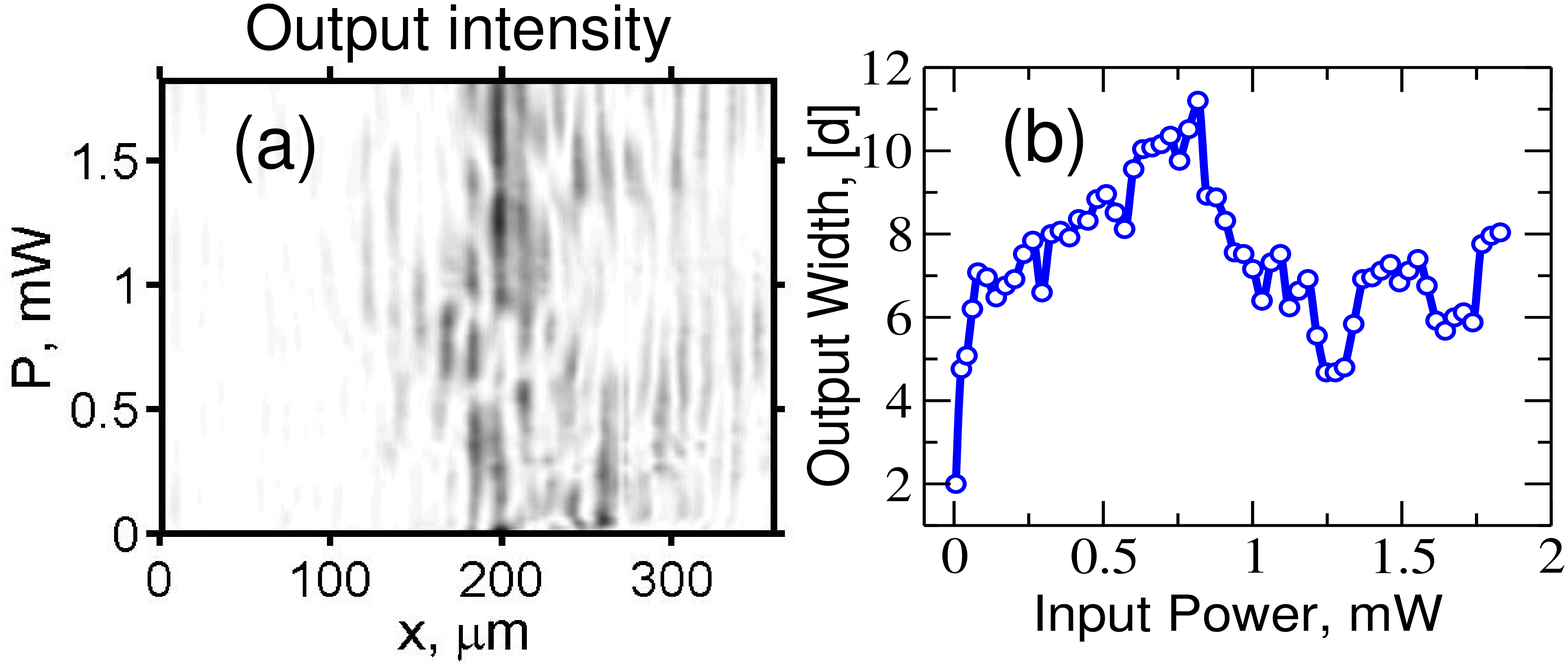}{fig5_Experiment_LiNbO3}{
(a)~Output beam profile as a function of the input power measured in curved LiNbO$_3$ waveguide array.
(b)~Output beam width vs. the input power. Wavelength is $\lambda = 532$~nm.
}

When only one waveguide is excited at the input, the light evolution should be fully equivalent for both positive and negative nonlinearities
in the framework of the tight-binding model~\cite{Longhi:2005-2137:OL, Longhi:2006-243901:PRL}.
In order to confirm that the effect of the nonlinear beam diffusion 
takes place in diffraction-managed waveguide arrays with both focusing and defocusing nonlinearity~\cite{Garanovich:2007-9547:OE}, 
we have also fabricated curved waveguide arrays by titanium indiffusion in a $50$~mm long X-cut mono-crystal LiNbO$_3$,
which exhibits {\em defocusing} photorefractive nonlinearity. Similar to the case of focusing nonlinearity in silica glass,
we have also observed that as the input power is increased, the self-collimation regime is destroyed, 
and the nonlinear beam diffusion takes place, after which the nonlinear beam self-trapping occurs at higher powers.
In Fig.~\rpict{fig5_Experiment_LiNbO3} an example of such a nonlinear dynamics
is shown for the curved LiNbO$_3$ waveguide array which consisted of the two sections each L$ = 25$~mm long,
with the waveguide spacing d$=14$~$\rm \mu$m and the bending amplitude A$=24.5$~$\rm \mu$m. The bending amplitude had opposite signs in
the two successive array segments in order to improve the symmetry of the output beam profiles, 
as suggested originally in ~\cite{Garanovich:2006-066609:PRE}. We have checked using numerical simulations
of the parabolic equations for the electromagnetic field that quite strong radiation which is visible in
Fig.~\rpict{fig5_Experiment_LiNbO3}(a) is not a result of the beam scattering in the curved waveguide array, 
but it is rather caused by the excitation of high order modes in the waveguides.

In conclusion, we have observed experimentally, for the first time, nonlinear beam diffusion and formation of diffraction-managed solitons in periodically-curved arrays of coupled optical waveguides with fully suppressed effective linear diffraction for both focusing and defocusing types of nonlinearity. These regimes are {\em fundamentally different} from the nonlinear beam self-focusing or self-defocusing in a bulk medium, or discrete self-trapping of light in arrays of straight waveguides. We also confirmed experimentally, that the critical power required for the formation of lattice solitons in curved waveguide arrays is several times higher than the soliton power in straight waveguide arrays. Our results
pave the way for the creation of functioning nonlinear devices for all-optical manipulation of light.

\bibliography{abbrev,papers,books,db_art_nlinDiffusionExp}

\end{sloppy}
\end{document}